\documentclass[letter]{aa}
\usepackage{graphicx}
\usepackage{txfonts}
\usepackage{lipsum}
\usepackage{subcaption}
\usepackage{lscape}
\usepackage{placeins}
\usepackage{color}
\usepackage{natbib,twoopt}
\usepackage[hyphenbreaks]{breakurl}
\usepackage[breaklinks]{hyperref}
\bibpunct{(}{)}{;}{a}{}{,}
\definecolor{cobalt}{rgb}{0.06, 0.2, 0.65}
\hypersetup{
  colorlinks,
  citecolor=cobalt,
  linkcolor=[rgb]{0.8, 0.2, 1.0},
  urlcolor=cobalt,
}
\newcommand{\be} {\begin{equation}}
\newcommand{\ee} {\end{equation}}
\newcommand{\bea}{\begin{eqnarray}}
\newcommand{\eea}{\end{eqnarray}}

\begin{document}

\title{A formation mechanism for narrow rings around minor bodies}
\author{C. Beaugé\inst{1}\inst{2}
   \and E. Gianuzzi\inst{2}
   \and N. Trógolo\inst{1}
   \and A.M. Leiva\inst{1}
   \and F.A. Zoppetti\inst{1}\inst{2}
   \and M. Cerioni\inst{2}
       }
\institute{Observatorio Astronómico, Universidad Nacional de Córdoba, Laprida
854, (X5000BGR) Córdoba, Argentina\\
          \and Instituto de Astronomía Teórica y Experimental (IATE),
UNC-CONICET, Laprida 854, (X5000BGR) Córdoba, Argentina\\
          }
\date{}
\abstract
% context heading (optional)
{The recent discovery of narrow rings around minor bodies has raised many
questions regarding their origin and current dynamics. Sharp ring boundaries
seem indicative of shepherding moonlets, but none have been found. All rings lie
close to spin-orbit resonances (SORs) with the central body, particularly the
1/3, even though it is not clear how these may be related. Furthermore, in at
least one case the location of the ring is exterior to the Roche radius, adding
to the striking differences with respect to giant planets.}
% aims heading (mandatory)
{We study the dynamical evolution of a particle disk around a minor body,
perturbed by the non-spherical component of the gravity field, particle
collisions, and spin changes of the central mass linked to angular momentum
conservation. By varying key parameters, we search for cases where the combined
effects may lead to resonance capture and orbital configurations similar to
those that have been observed.}
% methods heading (mandatory)
{We performed N-body simulations of massless particles orbiting a central
spherical body with a co-rotating mass anomaly. Collisions were modeled by 
adopting a simple radial damping force. Angular momentum conservation links the
body's spin to the disk’s orbital evolution. Since the gravitational effect of
the test particles is neglected, this back-reaction is introduced externally
assuming ad hoc spin-down rates and disk mass.}
% results heading (mandatory)
{Interaction between non-sphericity and collisions leads to the formation of a
narrow ring that slowly recedes from the central mass. Spin-down of the minor
planet shifts the SORs outward, enabling resonant capture. For suitable
parameters, a portion — or all — of the initial disk can become trapped in the
1/3 SOR with the mass anomaly, in dynamically stable low-eccentricity orbits.
Although the required disk mass is high ($\gtrsim 1 \%$ of the central body),
long-term collisional erosion could reduce it to values that are consistent with
observed ringlets.}
{}
\keywords{Celestial mechanics -- Planets and satellites: rings -- Minor planets
-- methods: numerical}
\maketitle

\section{Introduction} \label{sec:intro}

As of the writing of this paper, dense and narrow rings have been detected
around three minor bodies: the centaur object Chariklo
\citep{Braga-Ribas.etal.2014} and two dwarf planets, Haumea
\citep{Ortiz.etal.2017} and Quaoar \citep{Morgado.etal.2023}. Although these
objects exhibit significant diversity in size (roughly between $127$ and
$712$ km) and rotational period (between $4$ and $17$ hours), the rings show
surprising similarities. All are located very close to spin-orbit resonances
(SORs) with the central mass, mainly the 1/3, and they show no evidence of any
shepherding companions to explain the lack of radial diffusion.\footnote{A
second ring around Quaoar discovered by \cite{Pereira.etal.2023} seems to lie
close to the 5/7 SOR.} Even more baffling, the location of the main ring around
Quaoar lies outside the Roche radius, raising the question as to why such a
dense population of bodies did not accrete into a single satellite. The
existence of a small outer moon (Weywot) close to a 6/1 mean-motion resonance
(MMR) with the ring may increase the eccentricities and inhibit accretion;
however, there is little concrete evidence of such a scenario.

The proximity to resonances is particularly challenging and contrasts with
the dynamical characteristics of ring systems around giant planets. To date no
plausible mechanism has been presented to explain how SORs could have shaped
the structure of rings around minor bodies or how orbital evolution could have
led to such configurations. Resonance capture requires convergent migration,
while the outward spiraling of ring particles due to collisions and/or
diffusive effects gives rise to a divergent migration. These difficulties have
resulted in a certain consensus that perhaps there is no causal relation between
the observed population and SORs.

In this letter we explore the possibility that resonance capture in SORs
could have occurred as a result of the combination of three effects: (i)
gravitational perturbations from the non-spherical component of the central
mass, (ii) collisions in the proto-ring, and (iii) secular changes in the
rotational period of the minor body due to angular momentum conservation of the
full (body plus ring) dynamical system. While the roles of the first two
interactions have been analyzed in the past, the back-reaction onto the
central mass due to the orbital evolution of the proto-ring has not been
considered and will prove crucial. For adequate values of system parameters,
we will show how these interactions, particularly the spin-down of the
central mass, can lead to resonance capture in a 1/3 SOR of a significant part
of an initial particle disk and the formation of a stable narrow ring similar to
those observed. Our analysis, however, is more oriented toward presenting a
proof of concept of this mechanism and is not aimed at reproducing any particular
detail of the known ring systems around Centaurs, trans-neptunian objects
(TNOs), or dwarf planets.

\section{Numerical simulations}

\subsection{Physical scenario and dynamical interactions} \label{sec:sys}

Following previous studies \citep{Sicardy.etal.2019, Madeira.etal.2022,
Giuliatti.etal.2023}, we approximated the gravitational field of the
irregularly shaped central body of total mass $M_c$ by a spherical component of
mass $M_0$, physical radius $R_0$, and spin frequency $\Omega_0$, plus a
point-mass anomaly $M_1$, fixed on its surface.\footnote{The value of $M_1$
could be negative, correpsonding to a crater on the surface of the body.} We
note that $M_c = M_0 + M_1$. Given the oblateness parameters estimated for
Chariklo ($\epsilon \sim 0.2$), Haumea ($\epsilon \sim 0.61$), and Quaoar
($\epsilon \gtrsim 0.1$), we expected the ratio $\mu = M_1/M_0$ to be
considerable, even if quantitative values are beyond our possibilities. The
barycentric positions of each mass component, $M_k$, (with $k=0,1$) is denoted
by ${\pmb \rho_k}$, with modulus $|\, {\pmb \rho_0}| = R_0 M_1/M_c$ and $|\,
{\pmb \rho_1}| = R_0 M_0/M_c$, both with angular frequency $\Omega_0$.

We assumed a particle disk orbited the central mass and was the precursor of the
observed ring structure. Its formation and origin lies beyond the scope of this
work, but we presume it resulted from a partially disruptive collision of
the central mass with an external body or from the fragmentation of a primordial
satellite \citep[see, e.g.,][]{Melita.etal.2017}. We assumed the disk to be
coplanar with the equator of the massive body and to be radially bounded between
$r_{\rm min}$ and $r_{\rm max}$. Additionally, we supposed a certain surface
density profile $\Sigma_{\rm disk}(r)$, and a total mass $m_{\rm disk}$. Any
initial eccentricity distribution proved immaterial, rapidly reaching
equilibrium values (denoted as $e_{\rm eq}$, see next sub-section) in a few
orbital periods. The size distribution of the particles is relevant only when
modeling their collisional evolution.

The equations of motion of each disk particle may be written, in a barycentric
reference frame, as:
\be
\ddot {\bf r} = \ddot {\bf r}_{\rm grav} + \ddot {\bf r}_{\rm col}
= -\sum_{k=0}^1 {\cal G} M_k \frac{({\bf r} - {\pmb \rho_k})}{ \,\, | \, {\bf r}
- {\pmb \rho_k} \, |^3} -\eta \, n \, {\dot r} \; \hat{\bf r}
\label{}
\ee
where ${\cal G}$ is the gravitational constant. The dynamical system is thus
described by a circular restricted three-body problem, $\ddot {\bf r}_{\rm
grav}$, with non-Keplerian motion of the primaries, plus an external radial
friction term, $\ddot {\bf r}_{\rm col}$. We disregarded mutual gravitational
interactions between elements of the disk, treating them as massless particles.
The external force $\ddot {\bf r}_{\rm col}$ is a simple analytical prescription
that mimics the dynamical effects of mutual collisions between ring elements
\citep{Sicardy.etal.2019}. In this expression, $n$ is the mean motion of the
particle, and $\eta$ is a dimensionless coefficient related (among other
things) to the particle size and density profile at the location. We explored
values between $10^{-4}$ and $10^{-2}$.

Since the external force is radial, it will preserve the angular momentum of
each particle and thus of the disk as a whole. While this simplistic
collisional model cannot compete with fully interacting N-body simulations, it
does appear to reproduce many of its general features \citep{Salo.etal.2021}.

\begin{figure}[!ht]
\centering
\includegraphics[width=\hsize]{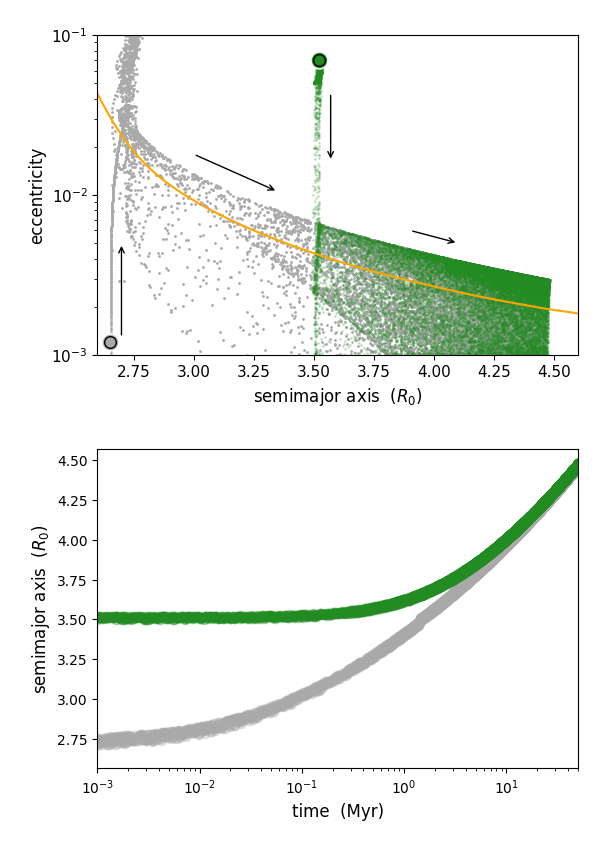}
\caption{Orbital evolution of two disk particles under the effects of
gravitational perturbations and collisional drag ($\eta = 10^{-3}$). The central
body was defined by $\mu = 0.05$, $R_0 = 115$ km and $\Omega_0/n_1 = 0.46$. {\bf
Top:} Eccentricity versus semimajor axis. Initial conditions are indicated in
filled circles, while arrows show the direction of flow. The equilibrium
eccentricity $e_{\rm eq}$, is highlighted in orange. {\bf Bottom:} Semimajor
axes as a function of time. Notice that the semimajor axes of the particles
converge over time.}
\label{fig1}
\end{figure}

\subsection{Equilibrium eccentricity and radial confinement}

Conservation of angular momentum by $\ddot {\bf r}_{\rm col}$ implies that
$a(1-e^2) = const$, and the orbits suffer a damping (proportional to $\eta$)
while exhibiting almost no change in the semimajor axis. Were it not for the
mass anomaly, an eccentric initial disk would end as a circular structure
analogous to the Saturn rings. However, the non-spherical component of the
central mass introduces an equilibrium eccentricity, $e_{\rm eq}$, around which
the particles oscillate (see Appendix \ref{appendix}). Under the effects of an
exterior non-conservative force such as our radial term, this stable solution
morphs into an attractor, and any initial eccentricity distribution tends toward
these values, hence the name "equilibrium eccentricity."

The top frame of Fig. \ref{fig1} shows the orbital evolution of two elements
of an extended disk. The initial conditions are shown with large filled circles,
while arrows indicate the flow. Regardless of the initial orbit chosen above
(green) or below (gray) $e_{\rm eq}$, the non-conservative force will rapidly
lead the eccentricity to oscillate around the equilibrium eccentricity
highlighted in orange. Once this occurs, the disk continues to evolve outward
(see Appendix \ref{appendix}), albeit at a much slower rate.

Perhaps the most interesting result of this section is shown in the bottom frame
of Fig. \ref{fig1}, where we compare the semimajor axes of the two disk 
particles over time. As $a$ grows, $e_{\rm eq}$ is strongly reduced. Even
though the eccentricity gradient also decreases, it is insufficient to
counteract the previous trend, and the rate of change of the semimajor axis
decreases. The resulting effect is an accumulation of the disk particles into an
increasingly narrow ring. The time necessary for this radial confinement depends
on several factors, including (i) the initial width and location of the disk;
(ii) the collision coefficient, $\eta$; (iii) the mass ratio, $\mu$; and (iv)
the spin rate of the central body, $\Omega_0$. All other things being equal,
this mechanism indicates that slow rotators are more efficient at converting
disks into rings (Appendix \ref{appendix}).

\subsection{Rotational spin-down and resonance capture}

Provided the confinement mechanism described above is more effective than any
radial diffusion, the end product will be a stable narrow ring without the need
of any shepherding satellite. As shown in the bottom frame of Fig. \ref{fig1},
the ring will continue to recede from the central body, albeit at an increasingly
slowler rate. The eccentricity of the ring particles oscillate around $e_{\rm
eq}$, taking values of the order of $e \sim 10^{-4}-10^{-2}$, depending on the
system parameters. The dynamics, however, are completely secular, and so far
there is no reference to an SOR. The fact that all observed rings around
minor planets favor SORs, particularly the 1/3, indicates that other dynamical
phenomena must be at play.

For resonance capture to occur, orbital migration must be convergent
\citep[e.g.,][]{Beauge.Cerioni.2022}. In the case of exterior spin-orbit
resonances, this implies that the frequency ratio $n/\Omega_0$ must increase
over time. Since the orbital dynamics described in the previous section leads to
an increase of the semimajor axis of the disk particles and consequently a
reduction in the orbital frequency $n$, convergent migration can only occur if
the spin rate, $\Omega_0$, were to also decrease at an even faster rate.

Although a secular variation in the body's spin seems unlikely, it is
nonetheless expected when assuming the conservation of the total angular
momentum of the system. More importantly, its magnitude may be estimated from
the back-reaction of the disk migration acting on the rotation of $M_c$.

Even if self-gravity was not considered in our numerical simulations, the disk
is expected to have a certain mass $m_{\rm disk}$, following a surface density
profile, $\Sigma_{\rm disk}(r)$. By representing this population by a set of $N$
particles, each of mass $m_j$, semimajor axis $a_j$, and eccentricity $e_j$,
the orbital angular momentum of the disk may then be expressed as:
\be
L_{\rm disk} = \sum_{j=1}^N m_j |{\bf r_j} \times {\bf \dot{r}_j}|
 = \sum_{j=1}^N m_j \sqrt{{\cal G} M_c a_j (1-e_j^2)}.
\label{ang3}
\ee
Assuming that the disk retains radial symmetry throughout its orbital evolution,
the center of mass of the complete system will coincide with that of the central
mass, $M_c$. Consequently, the total angular momentum may be written in
barycentric coordinates as:
\be
L_{\rm tot} = \Lambda \Omega_0 + L_{\rm disk}
\hspace*{0.3cm} ; \hspace*{0.3cm} {\rm where} \hspace*{0.3cm}
\Lambda = \frac{2}{5} M_0 R_0^2 + \sum_{k=0}^1 M_k \rho_k^2  ,
\label{ang2}
\ee
with $\Lambda=\Lambda(M_0,M_1,R_0)$, a function of the physical characteristics
of the central body and independent of its rotation. The first term stems from
the rotational motion of $M_0$, while the second is the contribution from the
translational motion of the central dipole. Invariance of $L_{\rm tot}$ implies
that any increase in the disk angular momentum due to outward migration will
lead to a decrease in the rotational frequency of the central body.

\begin{figure}[!ht]
\centering
\includegraphics[width=\hsize]{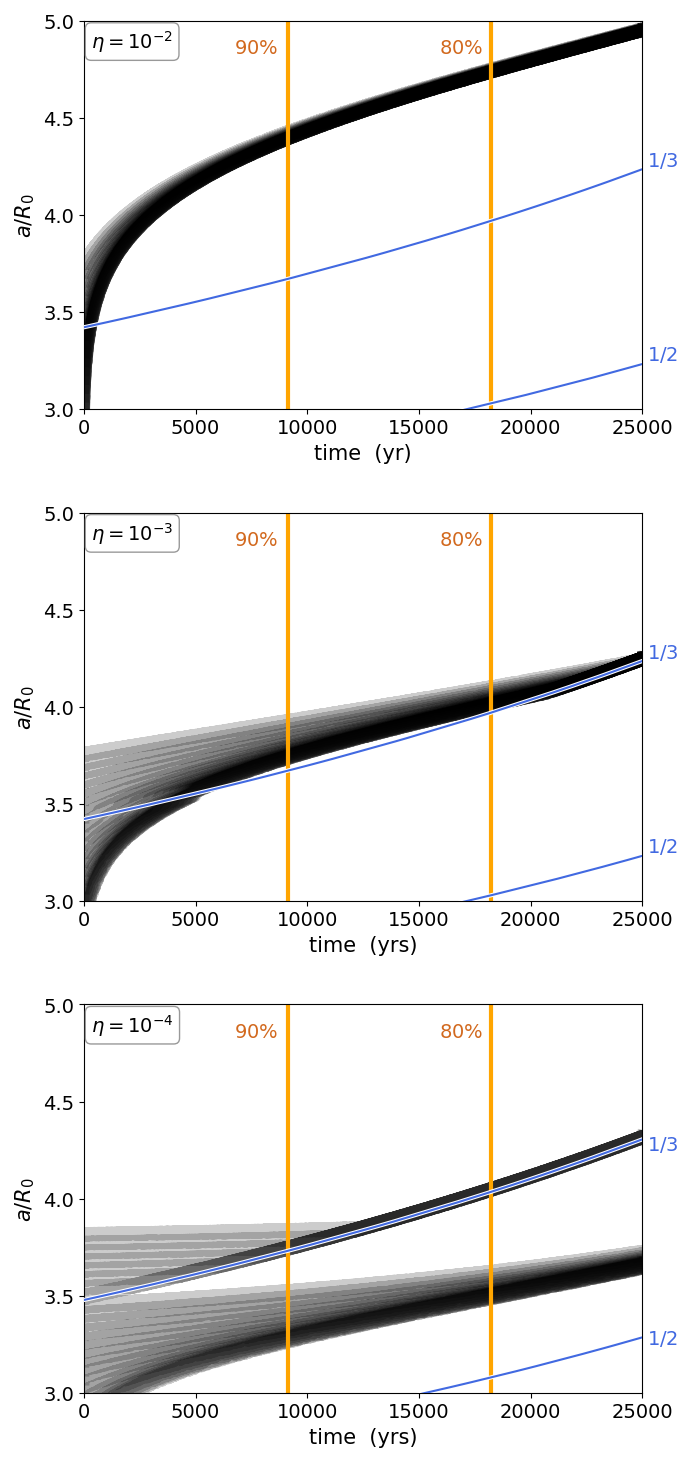}
\caption{Three sets of simulations of the dynamical evolution of particle disks
considering different values for the collisional coefficient, $\eta$. In all
cases, the mass anomaly was taken to be equal to $\mu = M_1/M_0 = 0.05$, while a
spin-down rate with e-folding time $\tau_s = 10^5$ years was adopted for the
rotational frequency. The blue curves show the semimajor axes associated with
the 1/3 and 1/2 SORs. The vertical lines indicate the times at which the
spin-down reached values of $\Omega_0$ equal to $90\%$ and $80\%$ of the initial
rotation frequency.}
\label{fig2}
\end{figure}

Lacking an explicit expression for $e_{\rm eq}(a)$, our analytical insights
could not go any further. However, we could perform numerical simulations
introducing ad hoc spin-down rates ${\dot \Omega}_0$, and study its effect on
the disk dynamics. Fig. \ref{fig2} presents three examples plotting the
semimajor axes of disk particles as a function of time. We chose a Chariklo-like
body for a central mass, with $M_c = 6.3\times10^{18}$ kg, a reference radius of
$R_0 = 115$ km, and a rotational period $P_{\rm rot} \equiv 2\pi/\Omega_0 =
7.004$ hs \citep{Leiva.etal.2017,Sicardy.etal.2019}. The non-sphericity was
modeled by a single mass anomaly with $\mu = 0.05$. The initial disks were
assumed to be circular with radial symmetry, between $r_{\rm min} = 2.5 R_0$
(slightly above the co-rotation radius) and $r_{\rm max} = 3.8 R_0$.

Each disk, consisting of 35 massless particles under the effects of
gravitational perturbations and radial drag, was integrated for $10^5$ years. We
also included an ad hoc spin-down of the central mass with a characteristic
timescale of $\tau_s = -{\Omega_0}/{\dot \Omega}_0$. The simulations shown in
Fig. \ref{fig2} used $\tau_s = 10^5$ years. Since the equilibrium eccentricity
is inversely proportional to $(\Omega_0-n)$, the spin-down helped fuel the
outward migration of the disk and greatly reduced the evolution timescale. More
importantly, variations in $\Omega_0$ moved the position of the SORs outward.

Each frame shows results for a different value of $\eta$. The inner initial
disk, corresponding to orbits with a semimajor axis of $a \lesssim 3.0 R_0$,
has large values of equilibrium eccentricities $e_{\rm eq}$, and thus suffers a
very fast radial migration, reaching separations of the order of $3 R_0$ in a
few decades. After this first stage, evolution continues more slowly. For $\eta
= 10^{-2}$ (top panel), the initial disk converges to a narrow ring in
timescales on the order of $T \sim 10^3$ years. However, the outward evolution
of the ring proved faster than that of the SORs, leading to divergent migration
and no resonance capture.

For $\eta = 10^{-3}$ (center panel of Fig. \ref{fig2}) the outcome is very
different. Even if the radial confinement takes longer ($T \sim 10^4$ years),
resonance capture is observed early in the process for initial conditions above
and below the location of the 1/3 SOR. The resonant dynamics is preserved
throughout the evolution of the narrow ring, with eccentricities displaying
short-period variations between $10^{-3}$ and  $10^{-2}$. The width of the ring
continues to decrease as it recedes from the central body and $M_0$ spins down,
reaching values analogous to those currently observed.

Finally, in the case of $\eta = 10^{-4}$ (bottom panel of Fig. \ref{fig2}) we
observed the formation of two rings. The first ring is narrow, captured again in
the 1/3 SOR. It formed a bit earlier than in the previous example and consists
of particles initially exterior to the resonance location. A second broader ring
is formed closer to the central body and is ultimately trapped in the 1/2 SOR.
However, this requires a spin-down of more than $50\%$ of the initial value, and
may thus be considered an unlikely result.

These simulations are illustrative and have been chosen to highlight the
different possible outcomes. While the mechanism leading to radial confinement
is extremely robust and was found in all simulations, resonance capture is more
sensitive. The SORs require a minimum width for trapping to take place, which
depends on the value of $\mu$. We found no resonant ring for $\mu < 10^{-3}$,
although the limit is a function of $\eta$. Similarly, convergent migration
requires minimum values for the spin-down rates, and such a condition was not
met for $\tau_s \gtrsim 10^6$ years. Within these ranges, outcomes analogous to
the bottom frame of Fig. \ref{fig2} were more frequent than that depicted in the
center panel. However, our exploration of the parameter space is far from
thorough.

Since our simulations are not self-consistent, we had to estimate the disk mass
necessary to generate the assumed spin-down rate used for Fig. \ref{fig2}. Once
again, we invoked the conservation of total angular momentum. Considering disk
elements of equal mass, we rewrote expression (\ref{ang2}) as:
\be
L_{\rm tot} = \Lambda \Omega_0 + m_{\rm disk} {\hat L}_{\rm disk}
\label{}
\ee
where ${\hat L}_{\rm disk}$ is the disk angular momentum per unit mass, a
value that can be calculated from the numerical simulations at any given time.
By calculating this quantity at two different times, say at $t=0$ and when the
spin was reduced by $20 \%$, we could calculate the values of $m_{\rm disk}$ that
preserves $L_{\rm tot}$.

The results from Fig. \ref{fig2} yield values for the total disk mass in the
range of $m_{\rm disk} \sim 8 \times 10^{-3} M_0$, showing only a weak
dependence with $\eta$. No significant changes were observed when varying the
time at which the disk angular momentum is evaluated, indicating a fairly robust
estimation. Although these values are much larger than the current estimations
\citep{Pan.Wu.2016}, collisional grinding within the dense ring could have
pulverized a large portion of the original mass later ejected by radiation
pressure \citep{Morgado.etal.2023}. This process could also aid in reducing the
spin-down and lead to a steady-state configuration.

\section{Conclusions}

In this work, we have proposed a dynamical model for the origin and evolution
of dense narrow rings around minor bodies, starting from an extended debris disk
and evolving toward dynamical structures similar to those observed in several
Centaurs and TNOs. Based on first principles, the observed radial confinement is
the natural outcome of the gravitational interactions with the central body and
a radial external force. Although we assumed a radial Stokes-like acceleration,
analogous results should be obtained for any non-conservative force resulting in
strong eccentricity damping.

The back-reaction on the central body's spin is also a crucial ingredient,
leading to convergent migration and (at least for some disk particles) resonance
capture. Coincidental with observed systems, the 1/3 SOR appears as the most
preferable commensurability. The radial confinement is maintained by resonant
dynamics and the eccentricity gradient, without the need of shepherding
satellite companions. The eccentricities remain low ($\sim 10^{-3}-10^{-2}$)
but are large enough to inhibit accretional collisions
\cite[e.g.,][]{Brilliantov.etal.2015}, and thus perhaps explain the existence of
rings beyond the Roche radius, as observed around Quaoar. Since no apsidal
alignment was detected, the rings would be observed as circular.

These promising results require debris disks that are more massive than
estimated for the observed rings. While we cannot rule out the limitations of
our numerical method and models for particle interactions (collisions and
self-gravity), this discrepancy could be partially solved by collisional
grinding over long timescales, which are particularly active once the localized
surface density increases as a consequence of radial confinement.

\begin{acknowledgements}
Simulations were carried out with the computing facilities hosted at IATE as
well as in the High Performance Computing Center of the Universidad Nacional de
Córdoba (CCAD-UNC). This research was funded by CONICET and Secyt/UNC.
\end{acknowledgements}

\bibliographystyle{aa}
%\bibliography{let_anillo}
\bibliography{aa56383-25}
\begin{appendix}

%\onecolumn
\section{The semi-secular and equilibrium eccentricities} \label{appendix}

In addition to the system parameters defined in \ref{sec:sys}, let $\theta_1$
denote the azimuthal angle of the position of $M_1$ in a given inertial
reference frame. In the two-body problem, where the motion of the perturber
$M_1$ with respect to the center of mass is keplerian, i.e. $\dot{\theta}_1 =
n_1 \equiv ({\cal G} M_c/R_0^3)^{1/2}$, the equilibrium eccentricity $e_{\rm
eq}$ lies close to the semi-secular forced eccentricity $e_{\rm ff}$, associated
to stable fixed points in the restricted three-body problem, averaged over the
orbital period of the perturber. For initial conditions exterior to the binary
$M_0+M_1$, an octupole-level expansion of the disturbing functions yields:
\be
e_{\rm ff} = \frac{3}{4} \frac{\mu}{(1+\mu)^2} \biggl( \frac{R_0}{a}
\biggr)^2 + \frac{45}{64}\frac{\mu(1+\mu^3)}{(1+\mu)^5}
\biggl(
\frac{R_0}{a} \biggr)^4 .
\label{eq2}
\ee
A simplified version, limited to the quadrupole level, was found by
\cite{Paardekooper.etal.2012}. However, given the proximity of the future ring
to $M_c$, an extension to octupole terms proves necessary. It is important to
note that $e_{\rm ff}$ is equivalent to the classical secular forced
eccentricity for exterior orbits ($a > R_0$) in the case where the perturber
lies in a circular orbit.

\begin{figure}[!ht]
\centering
\includegraphics[width=0.95\hsize]{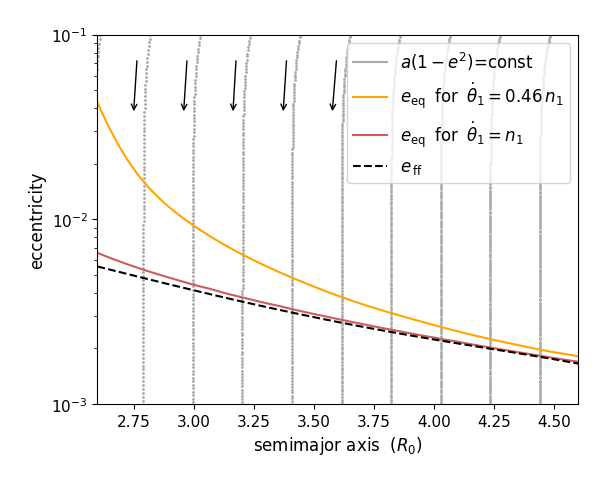}
\caption{Schematic view of the different interactions acting on the initial
particle disk. Curves of constant angular momentum (per unit mass) are shown in
gray, with arrows indicating evolutionary routes due to a radial drag term. Red
and orange curves highlight the equilibrium eccentricities of initial conditions
undergoing both collisions and gravitational perturbations (assuming $\mu =
0.05$ and $R_0 = 115$ km). Both differ in the assumed angular frequency
$\dot{\theta}_1$ of $M_1$ around the center of mass of $M_c$. The red curve
shows results assuming $\dot \theta_1$ equal to the keplerian mean-motion
($n_1$). Conversely, the orange curve was calculated considering a
sub-keplerian orbital frequency equal to the spin rate of $M_0$, i.e.
$\dot \theta_1 = \Omega_0 = 0.46 \, n_1$. For comparison, the black dashed curve
shows the semi-secular equilibrium eccentricity $e_{\rm ff}$, as given by
equation (\ref{eq2}).}
\label{figA1}
\end{figure}

The dashed black line in Fig. \ref{figA1} plots $e_{\rm ff}$ as function of
$a/R_0$, assuming a mass ratio $\mu = 0.05$. The exact values of the equilibrium
eccentricity, in the case of $M_1$ in keplerian orbit, are shown in red, and
were obtained from N-body simulations whose outputs were averaged over
timescales much larger than orbital period. As expected, $e_{\rm eq} \simeq
e_{\rm ff}$, and the discrepancy tends to zero for increasing semimajor axis.
Although the difference could be due to higher order terms in the Legendre
expansion of the disturbing function, its origin lies elsewhere and is much more
interesting.

We return to our minor body $M_c = M_0 + M_1$, where $M_1$ co-rotates
with $M_0$. In such a case, the motion of the perturber with respect to the
center of mass is sub-keplerian, with a corresponding angular ("orbital")
frequency equal to $\Omega_0 < n_1$. The orange curve in Fig. \ref{figA1}
shows the equilibrium eccentricity in this case, considering $\Omega_0 = 0.46
n_1$, a value similar to the current spin of Chariklo. The values of this new
equilibrium eccentricity are significantly higher than those obtained for
$\dot{\theta}_1 = n_1$ and the semi-secular limit $e_{\rm ff}$.

Numerical simulations, backed by perturbation theory, indicate that the change
in $e_{\rm eq}$ is related to the magnitude of short-period variations, which,
by construction, are not considered in the secular or semi-secular disturbing
functions. While a detailed model for the sub-keplerian equilibrium eccentricity
is outside the scope of this letter, its main features may be described by an
expression of the type: $e_{\rm eq}^2 - e_{\rm ff}^2 \propto
n/(k_1\Omega_0-k_2n)^2$, with $k_1$ and $k_2$ integers. Thus, slow-rotating
central bodies generate larger equilibrium eccentricities than fast rotators. As
the orbital frequency of $M_1$ increases, the difference with $e_{\rm ff}$
decreases, as observed comparing the red and dashed black curves.

For low eccentricities (say $\lesssim 10^{-2}$), the curves of constant angular
momentum are practically vertical en in $(a,e)$ plane and the dominant effect
is focused on the eccentricity. Thus, when the disk particle reaches the
equilibrium eccentricity, the non-conservative term has a non-zero component
along the gradient of $e_{\rm eq}(a)$. This eccentricity gradient then generates
a secular change in the semimajor axis toward larger values. The rate of change
in $a$ may be written as:
\be
\frac{da}{dt} \simeq \biggl( \frac{\partial e_{\rm eq}}{\partial a}
\biggr)^{-1} \left. \frac{de}{dt} \right|_{\rm col}
\simeq -\frac{1}{2} \eta \, n \, e_{\rm eq} \; \biggl( \frac{\partial e_{\rm
eq}}{\partial a} \biggr)^{-1} .
\label{dadt}
\ee
Since the equilibrium eccentricity gradient is always negative, all disk
particles recede from the central body over time.

\end{appendix}

\end{document}